\documentclass[multi]{cambridge6A}
\usepackage[numbers]{natbib}
\usepackage{rotating}
\usepackage{floatpag}
\rotfloatpagestyle{empty}
\usepackage{amsthm}
\usepackage{amsmath}
\usepackage{amssymb}
\usepackage{graphicx}
\usepackage{ubec}
\newcommand \beq{\begin{eqnarray}}
\newcommand \eeq{\end{eqnarray}}

\begin{document}

\author{C. J. Pethick}
\affiliation{The Niels Bohr International Academy, The Niels Bohr Institute, University of Copenhagen, Blegdamsvej 17, 
DK-2100 Copenhagen \O, Denmark  \par
NORDITA, KTH Royal Institute of Technology and Stockholm University, Roslagstullsbacken 23,
SE-10691 Stockholm, Sweden}

\author{Thomas Sch\"afer}
\affiliation{Department of Physics, North Carolina State University, \mbox{Raleigh, NC 27695-8202}, U.S.A.}
\author{A. Schwenk}

\affiliation{Institut f\"ur Kernphysik, Technische Universit\"at Darmstadt, \mbox{D-64289 Darmstadt,} Germany 
\par 
ExtreMe Matter Institute EMMI, GSI Helmholtzzentrum f\"ur Schwerionenforschung GmbH, D-64291 Darmstadt, Germany}

\chapter{Bose--Einstein condensates in neutron stars}
\vspace{10em}
\begin{abstract}
In the two decades since the appearance of the book ``Bose--Einstein Condensation'' in 1995, there have been a number of developments in our understanding of dense matter.  After a brief overview of neutron star structure and the Bose--Einstein condensed phases that have been proposed,  we describe selected topics, including neutron and proton pairing gaps, the physics of the inner crust of neutron stars, where a neutron fluid penetrates a lattice of nuclei, meson condensates, and pairing in dense quark matter. Especial emphasis is placed on basic physical effects and on connections to the physics of cold atomic gases.
\end{abstract}
\arabicfootnotes

\section{Introduction}

Neutron stars contain strongly interacting matter under extreme
conditions. The high densities, up to $\sim10^{15}$\,g\,cm$^{-3}$ in the interior, combined with the physics of
the strong interaction, which involves a rich spectrum of particles
and attractive interaction channels, can lead to the realisation of
various condensates in neutron stars. In this chapter, we give an
overview of these possibilities, ranging from pairing of nucleons, to
condensates of mesons, to quark pairing. After a description of the
structure of neutron stars, we provide an introduction to
the various condensates, focusing on new developments since the book
``Bose-Einstein Condensation'' in 1995 \cite{PSS:greenbook}.

We first discuss superfluid phases of neutrons and protons. These are rather well established and the present challenge is to make
reliable calculations of pairing gaps and critical temperatures. This
area has benefited significantly from the connections to cold atomic Fermi
gases with resonantly tuned strong interactions.  If hyperons are present in neutron stars, additional paired phases of hyperons are possible. At asymptotically high densities, the
interaction between quarks becomes weak and the pairing of quarks is
well established theoretically in this limit. We discuss how the
nature of the gluon-exchange interaction gives rise to special features for
quark pairing. At intermediate densities, the condensation of pions
and kaons is possible. Whether or not these phases exist in nature is an open question because of the difficulty of making reliable calculations of the effects of strong correlations:  this is true irrespective of whether one approaches the problem from low densities, using hadron degrees of freedom, or from high densities, where quark degrees of freedom are the natural choice. 
Finally, we discuss briefly some
observational consequences of condensation in neutron stars, including
the cooling of neutron stars, the impact on rotational behaviour and
on phenomena involving the neutron star crust.

\section{Neutron star structure}
Matter in a neutron star ranges in density from typical terrestrial values at the surface to greater than nuclear density at the centre.  In all but the outermost parts of the star, the thermal energy $k_B T$, where $k_B$ is the Boltzmann constant and $T$ the temperature, is low compared with characteristic  excitation energies of the system: the star is thus a low temperature system in which quantum effects play an important role.  Despite their name, neutron stars contain ingredients other than neutrons.  At the surface of the star, matter in its ground state consists of $^{56}_{26}$Fe nuclei, with an equal number of electrons to ensure charge neutrality:  thus, in the nucleus the number of neutrons, 30, is only slightly greater than that of protons, 26. The Fermi energy of the electrons increases rapidly with depth and it becomes energetically favourable for electrons to be captured by protons, thereby producing nuclei that are more neutron-rich.\footnote{For a review of the physics of the outer parts of neutron stars see \cite{PSS:cjpravenhall}.}  

\begin{figure}
\begin{center}
\includegraphics*[width=0.9\columnwidth]{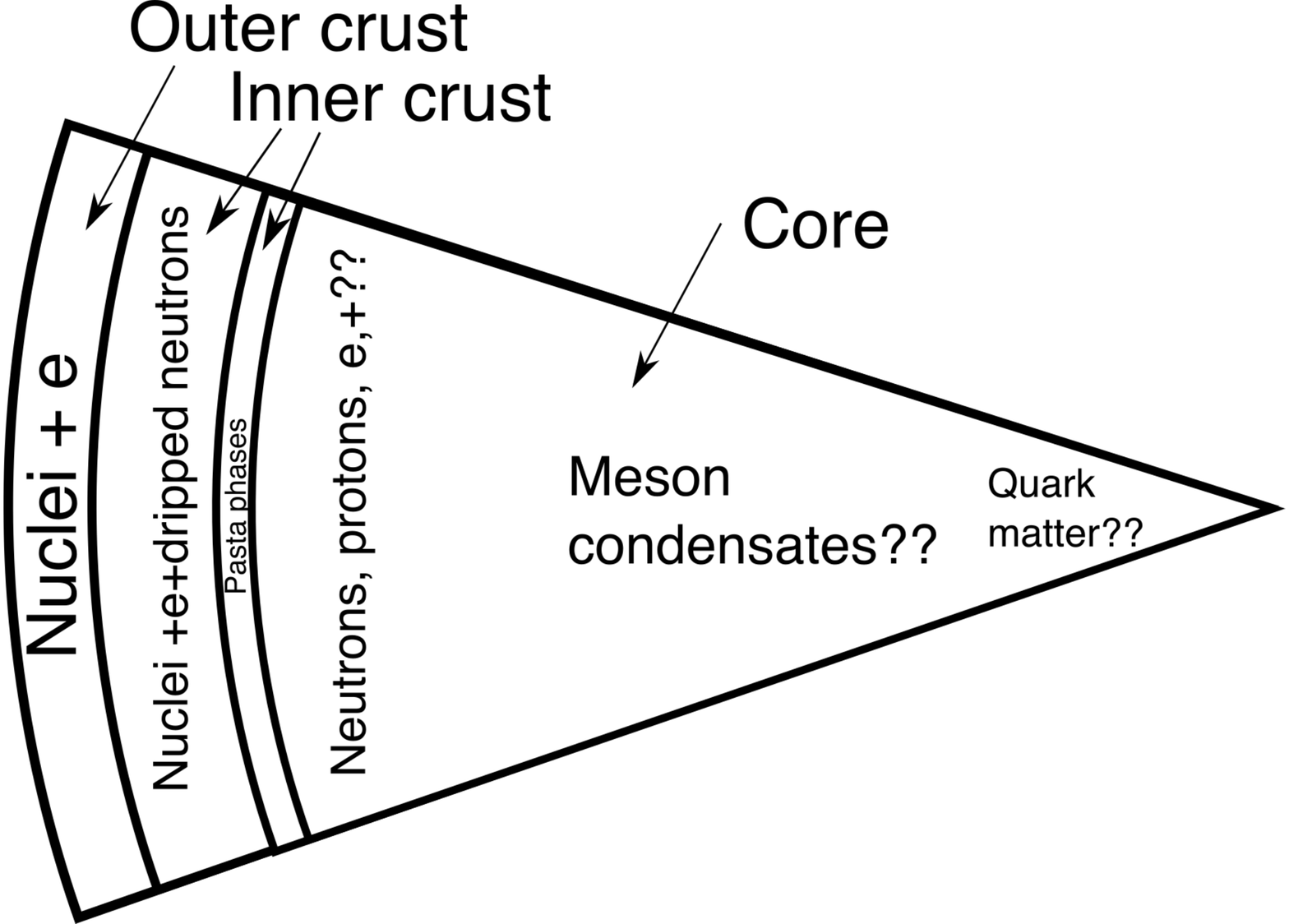}
\caption{Schematic picture of the phases encountered in a neutron star.} \label{PSS:fig:NSslice}
\end{center}
\end{figure}

At a density of around $4 \times 10^{11}$ g cm$^{-3}$, around one thousandth of nuclear density, the highest occupied neutron levels are no longer bound, a situation referred to as ``neutron drip''.  As a consequence,  at higher densities the lattice of nuclei is permeated by a neutron fluid.  With further increase in density, the density of this neutron fluid increases, and nuclei become even more neutron rich and occupy an increasing fraction of space.  
Nuclei merge to form a uniform fluid of neutrons and protons at a density of around one half of nuclear density, and the fraction of protons is $\sim 5\%$.
At densities just below that for the transition to a uniform medium, nuclei can form highly non-spherical shapes such as rods or sheets in what are referred to as ``pasta phases'' because of their resemblance to  spaghetti and lasagna.   
 At higher densities other constituents can appear: among these are muons, (which are present  when the electron chemical potential exceeds the muon restmass energy), hyperons and possible phases with deconfined quarks.  A schematic view of a slice of a neutron star is shown in Fig. \ref{PSS:fig:NSslice}.
The figure is not to scale.  For neutron stars in the mass range that can be observed, the radius is $\sim 12$ km \cite{PSS:HebelerLPS}, the outer crust is some hundreds of metres thick and the inner crust about one half of a kilometre.    

\section{Condensates in neutron stars}

We now give an overview of the various condensates that have been proposed.  These may be classified according to the baryon number,  $B$, of the condensed boson.\footnote{Irrespective of whether the condensed entity is a pair of fermionic excitations, as in a superconductor, or a bosonic excitation such as a meson, we shall refer to it simply as a ``boson''.}   
At the lower densities encountered in the inner crust and the outer part of core of the star, condensates of neutron pairs and of proton pairs, which have $B=2$, have been studied extensively.  These are analogous to condensates in conventional metallic superconductors and, in the case of pairs with nonzero orbital angular momentum, the superfluid phases of liquid $^3$He.  At supranuclear densities, condensates of pions and of kaons, which have $B=0$, have been proposed.    At even higher densities one expects matter to consist of deconfined quarks, and interactions between these can lead to condensates of pairs of quarks, which have $B=2/3$.  
\subsection{Pairing of nucleons}
\subsubsection{Neutron pairing}
\label{PSS:neutronpairing}

Even before the first identification of neutron stars in the cosmos and shortly after the formulation of the Bardeen--Cooper--Schrieffer (BCS) theory of superconductivity \cite{PSS:BCS}, Migdal in a side remark in a paper on superfluidity and the moment of inertia  of atomic nuclei commented that, in matter in the interior of neutron stars, neutrons would pair with a transition temperature of order 1 MeV ($10^{10}$ K) and that this would lead to interesting cosmological phenomena  \cite{PSS:Migdal}.
Estimates of transition temperatures for neutrons were made by Ginzburg and Kirzhnits \cite{PSS:GinzburgKirzhnits}, who also pointed out that the heat capacity of matter at temperatures significantly below the transition temperature would be reduced, thereby increasing the cooling rate of the star.  

Initially, pairing in the singlet, S-wave state ($^1$S$_0$) was considered.  However, on the basis of calculations of the pairing gap which used interactions deduced from nucleon--nucleon scattering data, Hoffberg et al.\  showed that, at densities in excess of roughly nuclear density, pairing in a spin-triplet state with unit orbital angular momentum would lead to a lower energy  \cite{PSS:Hoffberg}.  Explicitly, the state was found to be $^3$P$_2$, in which the orbital and spin angular momenta are aligned.\footnote{As a consequence of the tensor character of nucleon--nucleon interactions the state contains an admixture of the $^3$F$_2$ state, but for simplicity we shall refer to the state as $^3$P$_2$.}  This takes advantage of the fact that the nuclear spin--orbit interaction is attractive (in contrast to atomic physics where it is repulsive), thereby favouring  parallel spin and orbital angular momenta.  The state is closely related to the superfluid states of liquid $^3$He, which also have unit orbital angular momentum, but with the important difference that for $^3$He the spin-orbit coupling is much weaker, since it is due to the dipole--dipole interaction between the nuclear magnetic moments of the atoms.

\subsubsection{Proton pairing}  
\label{PSS:protonpairing}

In the outer core, the ratio of protons to neutrons is of order 5\%. 
In free space the interaction of two protons is closely equal to that between two neutrons.  This is due to the fact that the effects of electromagnetic interactions are small compared with the nuclear interactions, which are approximately invariant under rotations in isospin space. If the interaction between two protons in matter were the same as that in free space, this would imply that protons would be paired, with a gap  that depended on the  proton density in the same way as the neutron pairing gap depended on the neutron Fermi momentum.  In other words, the proton gap as a function of the total mass density would have the same basic form as that for neutrons, but shifted to $\sim 20$ times higher densities.    However, this picture is oversimplified, because the interaction between two protons is modified by the presence of the much denser neutron medium.    In addition, at these densities the contributions from many-body interactions are expected to be important. A reliable calculation of the proton pairing gap taking into account both these effects is an important open problem.

 \subsection{Meson condensates}

 \subsubsection{Pion condensation}
 The composition of matter in neutron stars was initially discussed in terms of models which treated the particles as being independent.  In particular,  Bahcall and Wolf \cite{PSS:BahcallWolf} pointed out that pions would be degenerate if their density were high enough.  If pions are treated as free particles, negative pions would appear in matter when the electron chemical potential became equal to the pion rest mass.   A macroscopic number of pions would then appear in the lowest energy state, forming a coherent state of the pion field. Because of interactions, the picture is more complicated, as described in detail by Baym and Campbell \cite{PSS:BaymCampbell}.  One of the key findings is that the most energetically favourable pion state is one with nonzero momentum.  This is due to the fact that the pion is a pseudoscalar Goldstone boson, and therefore the matrix element for absorption of a pion by a nucleon is proportional to $\boldsymbol\sigma\cdot {\bf q}$, where  $\boldsymbol\sigma$ is the nucleon spin operator and $\bf q$ the pion momentum. 
 For example, the interaction mixes  a negative pion  with proton-hole--neutron-particle states, thereby decreasing the energy of the pion.  Neutron and proton states with momenta differing by the pion momentum are mixed by interaction with the pion field and the elementary fermionic excitations are linear combinations of neutrons and protons.  The main obstacle to making reliable predictions of the threshold density for pion condensation is that it is expected to occur at supranuclear densities, where 
 central, tensor, and spin-orbit correlations as well as many-body forces are important.
 
 \subsubsection{Kaon condensation} 
 \label{PSS:sec:kaoncond}
  Because the mass of the strange quark is greater than that of up and down quarks, strange particles are not present in low density matter in equilibrium.  The mass of the strange quark is roughly 100 MeV, and therefore, when chemical potentials of constituents of matter (relative to their rest masses) are of order 100 MeV and greater, it is relevant to ask whether strange particles could be present.  
 In the normal state, there is a possibility of $\Sigma^-$ and $\Lambda^0$ hyperons being present.  Kaplan and Nelson \cite{PSS:KaplanNelson} proposed that, because of the attractive interaction between kaons and nucleons predicted by chiral theories,  condensates of kaons could appear in neutron stars.  In Ref.\ \cite{PSS:KaplanNelson} the calculations were made in the mean field approximation.  Subsequently, it was demonstrated that correlation effects would reduce the attraction between kaons and nucleons, thereby increasing the threshold density \cite{PSS:PandharipandeCJPThorsson}.    Estimating the threshold density for kaon condensation is challenging because of the paucity of experimental information of interactions of strange particles with other particles, in addition to the difficulties described for pion condensation.

 \subsubsection{Pairing of quarks}

 Up to this point, we have discussed possible Bose--Einstein condensed states of 
dense matter on the basis of a description in terms of hadronic degrees of freedom. In the regime of very high density it is more appropriate to use 
a Fermi gas of quarks as a starting point \cite{PSS:Collins:1974ky,PSS:Baym:1976yu}. 
It is difficult to quantify the meaning of ``high density'' in this context. 
In the case of high temperature and zero baryon density, numerical simulations 
of QCD on a space-time lattice show that a transition from hadronic matter 
to a quark gluon plasma takes place at a temperature of $T_c\simeq 170$ MeV. 
The transition is a smooth crossover, and the
equation of state of the plasma at temperatures $T\gtrsim 1.5 T_c$ can 
be described in terms of quark and gluon quasiparticles. At non-zero
baryon density lattice simulations cannot be carried out because
of the fermion sign problem, and as a result there is no reliable 
information on the phase diagram at high baryon density. Using the 
mean thermal momentum at $T_c$ as a guide for the Fermi momentum 
in quark matter near the critical quark chemical potential we 
get $\mu_c\simeq 500$ MeV, corresponding to a baryon density $\rho
\simeq 10\rho_0$, where $\rho_0 \approx 0.16$~fm$^{-3}$ is the nuclear saturation density. This is quite large, but there are indications
that the transition from quark matter to nuclear matter is 
smooth \cite{PSS:Schafer:1998ef}, so that calculations in the high-density limit may provide useful constraints on the behaviour of matter 
at moderate density. 

 The presence of a Fermi surface combined with the attractive interaction 
between quarks implies that the BCS mechanism will lead to Cooper pairing 
and superfluidity even if the gluon-mediated interaction is weak.  This was proposed by Ivanenko and Kurdgelaidze \cite{PSS:IvanenkoKurdgelaidze} even before the development of QCD.  Subsequently, following the understanding of asymptotic freedom, which implies that interactions between quarks  become weak at high densities, pairing between quarks of different flavours was considered in Refs.\ 
\cite{PSS:Barrois:1977xd,PSS:Bailin:1979nh}, see \cite{PSS:Alford:2007xm} for a review.    Interest in pairing of quarks revived about two decades ago with the appreciation that large pairing gaps can arise in states in which the flavours of quarks are correlated with their colours.

\section{Recent developments}
\label{PSS:Recent-developments}
In this section we describe some of the developments during the past two decades.
As we shall show, there are a number of points of contact between the physics of dense matter and that of ultracold atomic gases.

\subsection{Uniform neutron matter}
As described earlier, at densities above that for neutron drip, matter consists of a lattice of nuclei immersed in an electron gas with an interstitial fluid of neutrons, whose density increases from very low values just above neutron drip to ones approaching nuclear density at the inner edge of the crust.  For most of this density range, the nuclei occupy a small fraction of space and to a very good approximation the neutron fluid may be treated as being homogeneous.  This neutron fluid cannot be studied directly in the lab, and therefore its properties must be determined theoretically.  
The interaction between two neutrons in an S-wave state at low energy is described by the scattering length, 
which is $a_S \approx -18.5$~fm, and by an effective range $r_e \approx 2.7$~fm, which is relevant for the inner crust.
Since the interaction is attractive, a low-density neutron gas will  be paired in the $^1$S$_0$  state.    In what we shall refer to as the BCS approximation, one assumes that the interaction between two neutrons is the same as the interaction in free space, and the effect of the medium on the interaction  and on the single-particle properties is neglected.   One then finds that the neutron pairing gap in the dilute limit is given by 
\begin{equation}
\Delta_{\rm BCS}=\frac{8}{{\rm e}^2}E_F\exp{\left(-\frac{\pi}{2k_F |a_s|}\right)},
\end{equation} 
where $E_F=\hbar^2 k_F^2/2m_n$ is the Fermi energy, $k_F$ being the Fermi wave number and $m_n$ the neutron mass.

It came as something of a surprise to find that this is a poor approximation even in the limit of low densities.  This was well understood long ago by Gor'kov and Melik-Barkhudarov \cite{PSS:GorkovMB} in a paper that was largely unnoticed until attention was drawn to it after the experimental realisation of degenerate atomic Fermi gases \cite{PSS:HeiselbergPSV}.   They showed that in the low-density limit the gap is given by
\begin{equation}
\Delta=\left(\frac{2}{{\rm e}}\right)^{7/3}E_F\exp{\left(-\frac{\pi}{2k_F |a_s|}\right)},
\label{PSS:GorkovMB}
\end{equation} 
which is $(4{\rm e})^{-1/3}\approx 0.45$ of its value in the BCS approximation.   The suppression is readily understood in terms of the modification of the interaction between two neutrons by the presence of other neutrons.    At the lowest densities, the excitations in the medium are particle--hole pairs, which correspond to either density fluctuations or spin fluctuations.  Exchange of density fluctuations leads to an attractive interaction, as is well know in the case of conventional superconductors, where the density fluctuation is a lattice phonon.   However, exchange of spin fluctuations, i.e., particle--hole pairs with spin 1, gives a repulsive interaction.  For exchange of spin fluctuations with spin-projection $m_S=0$ the interaction between a neutron with spin up and one with spin down is positive, because the two neutrons exchanging the spin fluctuation have opposite spins and therefore couple to the spin fluctuation with opposite signs.  For excitations with $m_S=\pm1$, the interaction is again repulsive because, although the interaction is intrinsically attractive it corresponds to an exchange interaction in the pairing channel since it reverses the roles of the spin-up and spin-down neutrons undergoing pairing, thereby introducing an additional minus sign.

At higher densities, other many-body effects must be taken into account, and a variety of methods have been used to calculate the properties of neutron matter.  The best present-day techniques are  a family of Quantum Monte Carlo methods, which are accurate enough at lower densities that it is possible to calculate pairing gaps from the differences between the energies of systems with odd and even particle numbers (For a review of these developments, see \cite{PSS:Gezerlis}).  Thanks to these methods, coupled with the analytical results in the low-density limit,  the neutron gap is now well understood at densities less than about one tenth of nuclear density, where the effects of the two-body interactions beyond the S-wave contribution and of three-body interactions may be neglected.  The current challenge is to calculate neutron pairing gaps at higher densities, at which correlations become stronger and three-body interactions play an important role. (At saturation density, the leading four-body forces in chiral effective field theory have been shown to be very small, an order of magnitude smaller than 3N interactions.)

 To calculate gaps for non-S-wave pairing is more difficult, because these depend not on the pairing interaction averaged over angles between the initial and final momenta of the fermions undergoing scattering, as in the case of S-wave pairing, but on deviations of the pairing interaction from this average value.  For $^3$P$_2$ pairing, the role of correlations, in particular from induced spin--orbit interactions, has only been explored at low order in a many-body expansion \cite{PSS:SchwenkFriman};  these calculations led to very small $^3$P$_2$ gaps, but the possibilities that the $^3$P$_2$ gap vanishes or that pairing is stronger in some other channel are not excluded.

\subsection{Superfluid neutrons in the inner crust}
\label{PSS:innercrust}

In the inner crust, the neutron superfluid permeates a lattice of nuclei.  Low-frequency, long-wavelength phenomena may be described by a two-fluid model, the two fluids corresponding, loosely speaking, to the superfluid neutrons and the nuclei \cite{PSS:ChamelCJPReddy}. For a spatially uniform neutron fluid, at zero temperature Galilean invariance leads to the conclusion that the neutron superfluid density is equal to the total neutron density.  However, for neutrons in a lattice of nuclei, the neutron superfluid density is less than the total neutron density because of what in the language of quantum liquids is referred to as ``backflow''  and in the neutron star literature as ``entrainment''.  The two terms reflect different aspects of the problem, the first emphasising the disturbance of the neutron current density caused by the nuclei, the second the fact that part of the neutron density is locked to the nuclei and does not contribute to the superfluid flow.  

The neutron superfluid density is an important quantity in models of a number of phenomena observed in pulsars.  One is sudden speed-ups (glitches) of the rotation frequency of the pulsar, which in some models is attributed 
to superfluid neutrons weakly coupled to the crust and other normal parts of the star (for reviews of glitch phenomena see \cite{PSS:HaskellMelatos, PSS:Link}): here the moment of inertia  of the superfluid neutrons, which is directly related to the neutron superfluid density, plays a key role (see, e.g., \cite{PSS:Andersson,PSS:Chamel2013}).  Another is in explaining quasiperiodic oscillations observed in the X-ray afterglows of  giant flares from highly magnetised    
neutron stars, a possible model for which is oscillations of the crust of the neutron  star \cite{PSS:Duncan, PSS:SteinerWatts}.  The frequency of these modes is sensitive to the neutron superfluid density in the crust.

To the extent that the lattice is rigid, the system thus resembles a superfluid atomic Fermi gas in a three-dimensional optical lattice.  An important difference, however, is that the number of neutrons per unit cell  can be as high as $\sim 10^3$.  As a consequence of the large number of neutrons per unit cell, a corresponding large number of bands in the neutron band structure must be taken into account: this represents a considerable challenge.  If the effects of neutron pairing are weak, the neutron superfluid density is simply related to the response of a current in the normal state to a vector potential.  Chamel has performed mean-field (Hartree--Fock) calculations of the band structure of neutrons in the normal state and finds that the neutron superfluid density can be a factor of 10 or more smaller than the density of neutrons between nuclei, which one might expect to be a reasonable first estimate of the superfluid density \cite{PSS:chamel}.

An important open problem is to make improved calculations of the neutron superfluid density that take into account both band structure and pairing.  As an indication of the need to include both effects similtaneously, one may mention the fact that in the part of the inner crust where the neutron gap attains its largest values, the coherence length, which is the dimension of a neutron pair, is less than the lattice spacing.  The corresponding problem for fermionic atoms with resonant interactions in a one-dimensional optical lattice has been addressed in Ref.\ \cite{PSS:Watanabe1} and recently  Watanabe has extended these calculations to atoms with interactions that are not resonant \cite{PSS:Watanabe2}.  What is needed is an extension of this work to three-dimensional lattices and to higher numbers of atoms per unit cell.

\subsection{A dilute solution of protons in neutrons}

In the outer core of a neutron star, matter is expected to be a uniform fluid of neutrons with a $\sim 5\%$ admixture of protons (together with an equal admixture of electrons to maintain charge neutrality).  Since the protons are a minority component and the coupling of neutrons and protons is strong, interactions between protons in the medium are modified by the presence of the neutrons.  Dilute mixtures of atoms have been studied extensively, as have dilute solutions of $^3$He in liquid $^4$He, where solutions with a concentration of a few per cent have received particular attention.  

To illustrate how insights from cold gases may be exploited, let us consider the interaction of two protons with opposite spins induced by the presence of the nuclear medium.\footnote{The analogous problem of the interaction of minority fermions in the presence of a majority fermion component has previously been considered in the context of cold atomic gases \cite{PSS:MoraChevy, PSS:YuZoellnerCJP}.} Since the neutron medium has both spin- and  density degrees of freedom, the induced interaction has two contributions, one from exchange of density fluctuations (as in dilute solutions of $^3$He) and the other from exchange of spin fluctuations (see, e.g., Ref.\ \cite{PSS:BaldoSchulze}).   A new feature of the proton solutions compared with the  other dilute solutions is that tensor forces are important at the relevant densities, and the theory needs to be developed to include them.  In addition, the effects of three-body interactions, which are relatively unimportant in atomic gases play an important role at densities comparable to that of nuclei and above.

\subsection{Atomic analogue of pion condensation}

The production of cold gases of atoms with large magnetic moments has stimulated interest in creating atomic  systems in which there is a ``magnetic field condensation'' akin to the condensation of the pion field.  In this state the atoms exhibit a static spin-density wave. which is accompanied by a spin polarisation of the nucleons.  As an example, we mention the creation of cold, trapped dysprosium gases, which have magnetic moments roughly 10 times those of alkali atoms \cite{PSS:Lev1,PSS:Lev2}.  Viewed from the perspective of the particles,  the similarity between the two situations is clear: the dipole--dipole interaction between atoms and the one-pion-exchange (OPE) interaction between nucleons, both have a tensor structure.  

\begin{figure}
\begin{center}
\includegraphics*[width=0.7\columnwidth]{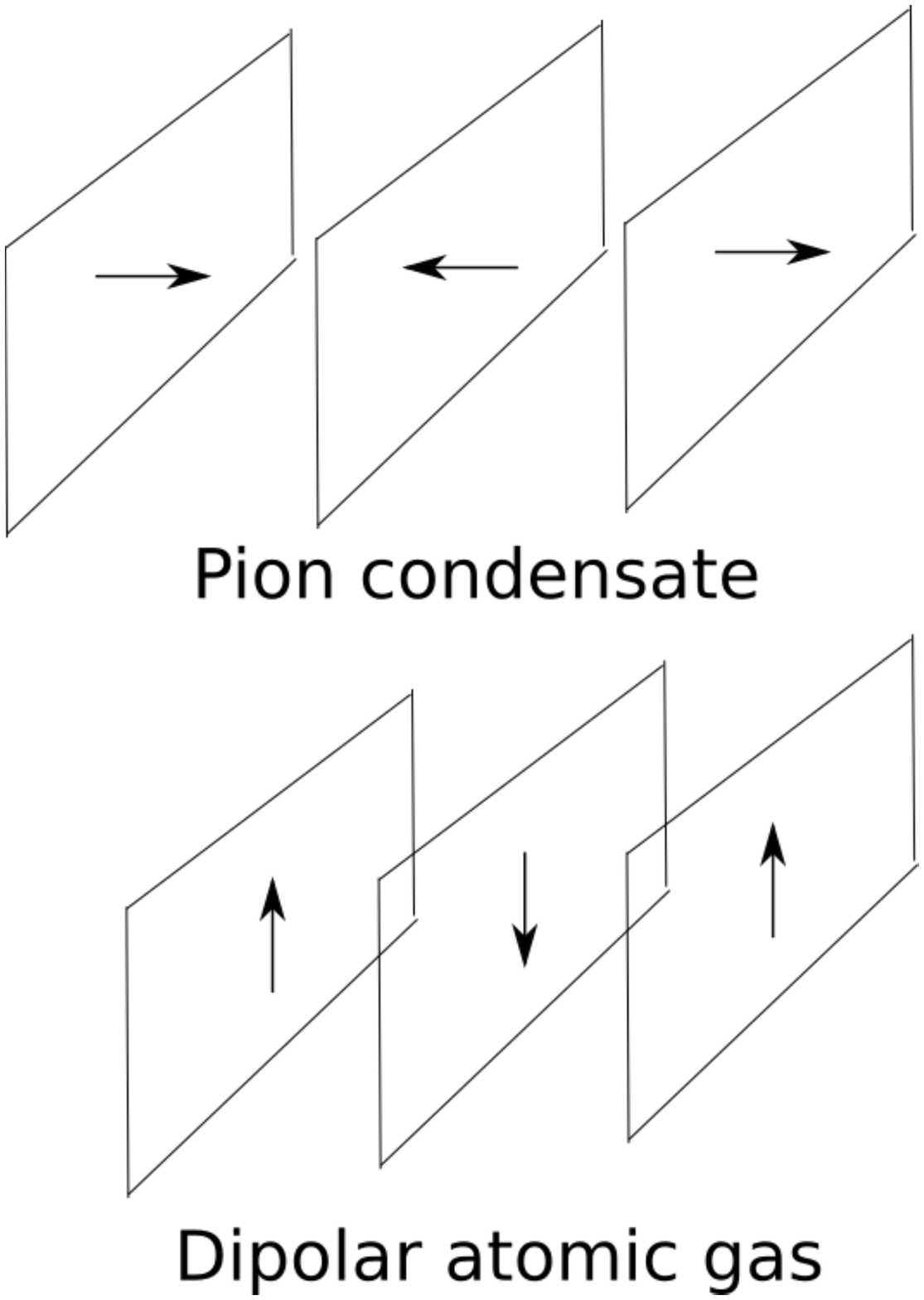}
\caption{Schematic picture of the neutron spin density wave in  a neutral pion condensed state (upper diagram) and in an atomic gas with dipolar interactions (lower diagram).  The magnetization density is uniform on the planes indicated and varies smoothly in between.  Due to nonlinear effects, the  spin-density wave will be accompanied by a density wave with a wave vector equal to twice that of the spin-density wave.  
For this reason, the phases are referred to as ``smectic" because of their resemblance to smectic liquid crystals.} \label{PSS:fig:AntiferroLayers4}
\end{center}
\end{figure}

An important difference between the two cases is the sign of the interaction: the electromagnetic dipole--dipole is negative for dipoles with the same orientation arranged head-to-tail, but positive for like dipoles lying side-by-side.  While the tensor force between a neutron and a proton in the isospin singlet channel has the same sign as for magnetic diples, the opposite is the case for the spins of neutrons interacting via the one-pion-exchange interaction. For a sufficiently strong dipolar interaction, uniform matter is unstable to formation of a spin-density wave.  For dipolar atoms in a one-dimensional structure, the local spin polarisation in the energetically favoured state  is perpendicular to the direction of the modulation \cite{PSS:Maeda}, while for neutrons interacting via the OPE interaction the polarisation is in the same direction as the modulation, as indicated schematically in Fig. \ref{PSS:fig:AntiferroLayers4}. If the dipolar interaction is weak compared with the central part of the interaction, a ferromagnetic state is predicted to be favourable.

The magnetic dipolar interaction differs from the OPE interaction in that it is long-range.  As a consequence, in finite geometry, such as a trapped atomic cloud, the equilibrium state will be influenced by the energy of the magnetic field outside the cloud.   As an example, consider matter which in bulk is predicted to be ferromagnetic: in a cloud the direction of the magnetisation will vary in direction in order to reduce the magnetic field energy outside the cloud, an effect analogous to formation of domains in a solid ferromagnet.  However, because of the absence of the lattice, the thickness of the domain walls is comparable to the size of a domain. 
Further work is required to determine in detail the structure of dipolar atomic gases, as a function of the strengths of the central and dipolar parts of the  interaction, the strength of the magnetic field, and the finite extent of the atomic cloud.  

\subsection{Quark matter}

\subsubsection{Normal state}

Even in the normal state, quark matter is an interesting system because it is an example of a marginal Fermi liquid:  the velocity of a quasiparticle with momentum $\bf p$ close in magnitude to the Fermi momentum, $p_F$, tends to zero as $p\to p_F$, whereas in a normal Fermi liquid the velocity tends to a constant.  The reason for this unusual behaviour is that, because gluons are massless, the gluon-exchange interaction is long-range.  The interaction is made up of colour-electric (longitudinal) and colour-magnetic (transverse) contributions in essentially the same way as for  the electric and magnetic contributions to the interaction between two electrical charges.   
The coupling of gluons to quarks is governed
by a coupling constant $g$ that depends on the momentum transfer. The sign of $g$, like that of the charge of the electron, is a matter of convention and we take it to be positive. The dimensionless quantity $g^2/\hbar c$ is the QCD analogue of the fine structure constant $\alpha=e^2/\hbar c$ in QED.
Asymptotic freedom implies that the typical interaction between quarks 
decreases as the density and the Fermi momentum increase. 
 If one neglects the effects of the medium on the gluons, the quark self energy due to emission and absorption of a gluon diverges.  In a medium, the electric part of the interaction is cut off at wave numbers less than the  screening wave number $k_D\sim g p_F$ by Debye screening.\footnote{To avoid making formulae cumbersome, in the remainder of this section we use units in which $c$, $\hbar$ and the Boltzmann constant $k_B$ are equal to unity.}  However, because quarks do not possess a magnetic charge, magnetic fields are not screened at zero frequency, but at nonzero frequency $\omega$, magnetic interactions are cut off at wave numbers less than   $(g^2|\omega| p_F^2)^{1/3}$ by Landau damping. In more technical terms, the transverse gluon propagator has the form
\cite{PSS:Baym:1990uj} 
\begin{equation}
\label{PSS:Pethick_D_ij}
D_{ij}(\omega,{\bf k}) = \frac{\delta_{ij}-\hat{k}_i\hat{k}_j}
 {\omega^2-c^2k^2+i\pi k_D^2 \omega/k}\, 
\end{equation}
for $|\omega|\leq k$.
In the physics of normal metals, the analogous phenomenon for electrodynamics is known as the anomalous skin effect: when the mean free path of an electron becomes larger than the wavelength of an applied magnetic field, the length relevant in determining the response is the wavelength rather than the mean free path.    

So far we have concentrated on the interaction between quarks, and have assumed that near the Fermi surface in the normal state the spectrum is linear in $p-p_F$.  However, from a one-loop calculation  using Eq.~(\ref{PSS:Pethick_D_ij}) for the one-gluon transverse propagator one finds that the quark propagator behaves as
\begin{equation}
\label{PSS:Pethick_S_ab}
 S(\nu,\vec{p}) = \frac{Z_F}
   {\nu - v_F(p-p_F)}\, , 
\end{equation}
with $Z_F\sim v_F \sim [g^2\log(k_D/|\nu|)]^{-1}$. This implies that the 
quasiparticle wave function and velocity vanish on the would-be
Fermi surface.\footnote{The analogous effect for the electron gas, in that case due to exchange of transverse {\it photons}, was investigated long ago in Ref.\ \cite{PSS:Holstein}.  However, in a nonrelativistic electron gas, the effects of transverse photons are of order $v_F^2/c^2$ smaller than those due to the Coulomb interaction and are not observable experimentally. } One might suspect that this breakdown of Fermi liquid theory would severely affect estimates of gaps in superfluid phases, but this turns out not to be the case, as we shall describe below.

\subsubsection{Order parameter}

 We turn now to the pairing interaction between quarks.  Since quarks are nearly massless the system is relativistic and
$p_F\simeq \mu_q$, where $\mu_q\simeq \mu_B/3$ is the quark chemical 
potential and $\mu_B$ the baryon chemical potential. In relativistic theory, fermions of a particular species are described by four-component (Dirac) fields rather than the two-component spinors corresponding to the two spin components in nonrelativistic theory. The fields describing quarks, $q^a_{\alpha f}$, are thus labelled by a Dirac index $\alpha$, in  addition to a colour index
$a$ (red, blue, or green) and a flavour index $f$ (up,
down, or strange).\footnote{Quarks of other flavours have much higher masses and would appear only at densities considerably higher than those anticipated to occur in neutron stars.} The interaction
between quarks is mediated by gluon fields $A_\mu^c$, which couple
to the colour charges of the quarks. Here, $c=1,\ldots,8$ labels
traceless hermitean $3\times 3$ matrices that act on the colour 
indices of the quarks. The one-gluon 
exchange interaction is attractive in the colour-antisymmetric quark--quark 
channel, and therefore, on the basis of the analogy with the nonrelativistic problem, one would expect quark matter to be unstable to formation of a state with quark pairs (diquarks).
 These pairs may be viewed as building blocks in the formation 
of baryons: diquarks $D_a=\epsilon_{abc}q^bq^c$ can bind with quarks
$q^a$ to form colour neutral baryons $B=D_aq^a$. 

 The order parameter is of the form
\begin{equation}
 \langle q^a_{\alpha f} (C\Gamma)^{\alpha\beta} q^b_{\beta g}\rangle 
  = \phi^{ab}_{fg}  \, , 
\label{orderparameter}
\end{equation} 
where $C$ is the charge conjugation matrix and $\Gamma$ is a Dirac
matrix.\footnote{This expression may be regarded as the generalisation to relativistic particles of the Nambu formalism for superconductivity.  There one works with (two-component) Pauli spinors and the order parameter is give in terms of spinor field operators $\psi_\sigma$ by $\langle \psi_\sigma{\sigma}_{i}\psi_{\sigma'} \rangle$, where $\sigma_i$ is a Pauli matrix. For example, with $\Gamma=\gamma_5$ the order parameter (\ref{orderparameter}) corresponds to the relativistic way of writing down singlet pairs, since the non-relativistic limit of $C\gamma_5$ is $\sigma_2$.} 
There are many possible channels, and the ground state will 
depend on the flavour composition and the strength of the interaction. 
At very high density matter is approximately flavour symmetric, and 
asymptotic freedom implies that the ground state can be determined 
in weak coupling QCD. One finds \cite{PSS:Alford:1998mk, PSS:Schafer:1999fe}
\begin{equation}
\langle q^a_{\alpha f} (C\gamma_5)^{\alpha\beta} q^b_{\beta g}\rangle  
  =  \phi \left(\delta^a_f\delta^b_g - \delta^a_g\delta^b_f\right) \, , 
\end{equation} 
which is called the colour--flavour locked (CFL) state.\footnote{This is the simplest choice for the order parameter and corresponds to a particular gauge choice.  A more general form may be obtained by performing a unitary transformation of the colour variables, which will leave unaltered variables that can be measured experimentally.} The order parameter 
is antisymmetric in colour, as required by the interaction, and 
antisymmetric in both flavour and spin. This implies that 
the quark pairs form spin singlets. As the density is lowered, the 
larger mass of the strange quark becomes more important, flavour 
symmetry is broken, and states with a smaller spin--flavour symmetry
may appear. 

 The colour--flavour locked state spontaneously breaks the chiral
symmetry of the QCD Lagrangian, and exhibits low-energy excitations
with the quantum numbers of pions and kaons. This can be seen as 
follows: The relation $C\gamma_5=C(P_L-P_R)$, where $P_{L,R}$ are 
projectors on left- and right-handedness, implies a fixed phase relation 
between the left- and right-handed components of the diquark condensate.
The colour and flavour orientations of the these condensates can be 
characterized by the matrices 
\begin{equation}
 X^a_f=\epsilon^{abc}\epsilon_{fgh}
   \langle (q_L)^b_g(q_L)^c_h \rangle , \hspace{0.5cm}
 Y^a_f=\epsilon^{abc}\epsilon_{fgh}
   \langle (q_R)^b_g(q_R)^c_h \rangle\ , 
\end{equation}
where the $\epsilon$-tensors take into account the antisymmetry
of the CFL state under exchange of colour and of flavour. The quantities  $X^a_f$ and $Y^a_f$ depend on the gauge choice but one can define a gauge invariant 
order parameter for chiral symmetry breaking by the relation $\Sigma_f^{\;g} = 
X_f^a(Y^\dagger)_a^g$.  

The CFL ground state corresponds to $\Sigma
\sim {\bf 1}$, and low energy modes with the quantum numbers of pions and 
kaons are described by oscillations of $\Sigma$ in the pion direction, $\Sigma\sim 
\lambda^{1,2,3}$, or the kaon direction, $\Sigma\sim \lambda^{4-7}$. Here $\lambda^a$ 
are the Gell-Mann matrices, a generalisation of the isospin matrices to 
flavour $SU(3)$. 

 In response to quark masses and lepton chemical potentials the CFL 
ground state can become polarised in the pion or kaon directions in flavour space. 
A non-zero 
strange quark mass, for example, favours a ground state in which 
there are fewer strange quarks than up and down quarks. This can be 
realized by a neutral kaon condensate $\Sigma\sim \exp(i\alpha
\lambda^{6})$ \cite{PSS:Bedaque:2001je}. This is a homogeneous condensate, 
analogous to the kaon condensate discussed from the point of view of hadronic matter in Sec.\ \ref{PSS:sec:kaoncond}. For larger strange quark masses or lower densities, 
interactions between kaons and gapless fermion modes can lead to the 
formation of a standing wave kaon condensate \cite{PSS:Schafer:2005ym}. 
This phase is analogous to the pion condensed state discussed above, 
and to standing wave ground states in Bose--Einstein condensed atomic gases, 
see \cite{PSS:Son:2005qx,PSS:Radzihovsky:2010}. Fermion quasiparticles 
in these systems are quarks surrounded by a diquark polarization
cloud \cite{PSS:Kryjevski:2004jw}, similar to the state $B=D_aq^a$ described 
above. As the strength of the interaction increases, these states
can continuously evolve into tightly bound baryons. A similar 
crossover can be studied in cold atomic gases of three fermionic
species \cite{PSS:Rapp:2006}.

\subsubsection{Gaps}

Superfluid gaps in quark matter have a very different dependence on coupling compared with dilute atomic gases.  Since the gluon exchange interaction is of order $g^2$, on the basis of the calculations for a dilute atomic gas, Eq.\ (\ref{PSS:GorkovMB}), one might have expected $\Delta \sim \mu_q e^{-{\rm const.}/g^2}$.
In fact, for weak coupling one finds  \cite{PSS:Son:1998uk, PSS:Schafer:1999jg,PSS:Pisarski:1999tv,PSS:Brown:1999aq,PSS:Schafer:1999fe}
\begin{equation}
\label{PSS:Pethick_gap_CFL}
 \Delta \simeq C_1\frac{\mu_q}{g^5} 
  \exp{\left(-C_2/g\right)}  \, ,
\end{equation}
where 
\beq
C_1 =\frac{512\pi^4}{2^{1/3}} \left(\frac{2}{3}\right)^{5/2} e^{-{(\pi^2+4)}/{8}},\,\,\,{\rm and}
\,\,\,
C_2=\frac{3\pi^2}{\sqrt{2}}.
\eeq
The reason for the difference lies in the long-range character of the gluon exchange interaction, as discussed above.            
Transverse gluons do not upset the basic mechanism of the 
BCS instability, but they do modify the magnitude of the gap. The gap 
equation has two logarithmic infrared divergences, one related to the 
BCS mechanism and one caused by the unscreened  transverse gluon exchange. 
The coefficient of the $1/g$ term in the exponent is determined by
transverse gluon exchange, and the exponential term in $C_1$ is related
to the wave function renormalisation $Z_F$ discussed above.  As one sees, the fact that, in the normal state,  quark matter is a marginal Fermi liquid has only a modest effect on the gap.  The 
pre-exponential factor is set by electric gluon exchanges, and is sensitive 
to the symmetries of the order parameter. For $\mu_q\simeq 500$ MeV 
the gap is of order $10$--$20$ MeV. 

\subsubsection{Superfluidity and superconductivity}
In the CFL phase, the densities of u-, s-, and d quarks are equal and,  consequently,  in bulk matter there is no need for other particles, such as electrons, to ensure electrical neutrality.    The CFL phase is a superfluid but an electrical insulator.  Heuristically, this may be understood  as being a consequence of the fact that the sum of the charges of the three flavours of quarks vanishes.\footnote{This argument is an oversimplification, as is explained in   \cite[Sec. II A.3]{PSS:Alford:2007xm}.} 

When the flavour symmetry is broken, the net charge density of the quarks is nonzero, and this is compensated by a background of electrons.  In this case, matter becomes a charged superfluid, and is an electrical superconductor.

\section{Observational considerations}

We now comment briefly on some observational consequences of Bose--Einstein condensation in neutron stars.  Models of glitches and X-ray flares that involve superfluid flow have been mentioned in Sec.\ref{PSS:innercrust}.  A more extensive account of observational effects may be found in Ref.\ \cite{PSS:PageLattimer}.

\subsection{Neutron star structure}
The mass--radius relation for neutron stars depends on the equation of state of matter, which is affected by  Bose--Einstein condensation.  In the case of pairing of nucleons and quarks, the predicted gaps are small compared with Fermi energies and, consequently, pairing has only a small effect on structure.  For meson condensates, the picture could be very different because, if it were energetically  favourable to create a pion or kaon condensate is, the energy would be reduced compared with that of the system with no condensate, and matter will become softer.   This would lead to a lower maximum mass of a neutron star.   The observation within the past few years of two neutron stars with masses approximately twice that of the sun with small error bars  \cite{PSS:Demorest, PSS:Antoniadis} is most simply understood in terms of an equation of state that is similar to what is expected for models based on nucleon degrees of freedom, without significant softening due to formation of a meson condensate.    

\subsection{Neutron star cooling}

For the first $10^5$-$10^6$ years after formation, a neutron star cools primarily by neutrino emission from its core.  In favourable cases, emission of X-rays from the surface of the star can be detected, and the temperature of the core deduced.  Since the rate of neutrino emission in dense matter depends on the microscopic properties of the matter, observations of cooling therefore have potential for providing information about the stellar interior (for a review see \cite{PSS:YakovlevCJP}).  The path from the microscopic properties of dense matter to observed cooling curves has many steps, but the past two decades have seen a steady increase in sophistication of the modelling and the precision of observations.

As an illustration of the potential such observations have for pinning down the properties of dense matter we give a recent example.   Pairing of nucleons reduces neutrino emission rates at temperatures well below the transition temperature because of the reduction in the number of thermal excitations.   However, just below the transition temperature, emission of neutrino--antineutrino pairs can occur by annihilation of two thermal excitations in the superfluid. In the normal state the process is forbidden by conservation of baryon number but in the superfluid, excitations are linear combinations of normal-state particles and holes and the process is allowed.  This process leads to accelerated cooling for a limited temperature range.  The reported detection of changes over a period of 10 years in the surface temperature of the neutron star produced in the Cassiopeia A supernova approximately 330 years ago has been interpreted as being due to neutrons in the core undergoing a transition to a state with $^3$P$_2$ pairing \cite{PSS:PageCasA, PSS:YakovlevCasA}.  While the analysis of the observations has been called into question, this case brings out the promise that such observations have for shedding light on pairing in neutron star cores.

\section{Concluding remarks}

The study of dense matter in neutron stars has profited from work on terrestrial quantum liquids and ultracold gases.  In addition, the studies of dense matter have provided inspiration for experiments in the laboratory.   In the future one may expect this synergy to continue.  

Within the next few years one may expect significant improvements in calculations of properties of matter at densities of order nuclear matter density by the application of chiral effective field theories \cite{PSS:Epelbaum:2008ga}.    Chiral effective field theory, which incorporates the symmetries of quantum chromodynamics, provides a systematic framework for treating interactions between nucleons and, when extended to strange particles, hyperons. Using a momentum expansion scheme, the theory includes the long-range interactions due to the exchange of pions (and kaons) explicitly and general contact interactions for the shorter-range parts. Moreover, it makes predictions for consistent three- and higher-body interactions. 

There are also aspects of superfluidity in neutron stars that are relatively little studied, among which one may mention superconductivity of protons in the pasta phases.  The appreciation of the fact that defects can occur in these phases and thereby lead to multiply connected structures \cite{PSS:Horowitz} implies that the proton superfluid may exhibit topologically nontrivial structures (flux lines) that are trapped by the defects.

\section{Acknowledgments}

We are grateful to Dmitry Kobyakov for helpful conversations and to
Gentaro Watanabe for communicating to us his results for the
superfluid density of an atomic Fermi gas in a one-dimensional optical
lattice. This work was supported in part by the NewCompStar network,
COST Action MP1304, the US Department of Energy grant 
DE-FG02-03ER41260, and by the ERC Grant No. 307986 STRONGINT.

\bibliography{PSS_References}
\bibliographystyle{cambridgeauthordate}

\end{document}